\documentstyle[12pt,twoside,fleqn,espcrc1]{article}

\newcommand{\AmS}{{\protect\the\textfont2 
A\kern-.1667em\lower.5ex\hbox{M}\kern-.125emS}}
\title{Fourier spectrum analysis of the new solar neutrino capture
rate data for the Homestake experiment}
\author{H.J. Haubold\address{Outer Space Office, United Nations, 
Vienna International Centre,\\ P.O. Box 500, 1400 Vienna, Austria}}
\begin{document}
\maketitle
\begin{abstract}
The paper provides results of the Fourier spectrum analysis of the new 
Ar-37 production rate data of the Homestake solar neutrino experiment and 
compares them with results for earlier data, revealing the harmonic 
content in the Ar-37 production in the Homestake experiment.
\end{abstract}

\section{THE SOLAR NEUTRINO PROBLEM}

All four solar neutrino experiments (Homestake, Kamiokande,
GALLEX, SAGE) observe a deficit of solar neutrinos compared to
the predictions of the Standard Solar Model. Additionally, any two of
the three classes of solar neutrino experiments indicate that the
largest suppression is in the middle of the solar neutrino
spectrum (the Be-7 line and the lower part of the B-8
spectrum)\cite{Kirs95}. Among these four experiments the Homestake
experiment is taking data for almost 25 years \cite{Davi92,Davi96}. The
reliability of the radiochemical method for detecting solar
neutrinos has been tested recently by the GALLEX experiment   
\cite{Kirs95}.

\section{IS THE SOLAR NEUTRINO FLUX CONSTANT?}

All efforts to resolve the solar neutrino problem by improving
solar, nuclear,and neutrino physics have not proven successful
so far\cite{Bahc89}. This may also mean that the average solar neutrino
flux extracted from the four experiments may not be the proper
quantity to explain the production of neutrinos in the
gravitationally stabilized solar fusion reactor \cite{Haub95,Gran96}. 

This conjecture is supported by the fact that the pattern of
the detected solar neutrino flux in the four solar neutrino
experiments hints at possible variation of the solar neutrino
flux over time, which has been generated interest to look at
time dependent phenomena \cite{Bahc89,Dick78,Rabe89,Haub91,Mcnu95} 
related to solar physics and neutrino physics. 

\section{FOURIER SPECTRUM OF THE HOMESTAKE DATA}

Figs. 1 and 2 show the Ar-37 production rate data which Davis and 
collaborators \cite{Davi92,Davi96} have acquired and analyzed in more than 
25 years. For the analysis of the radioactive decay of argon with a small 
number of counts the method of maximum likelihood was employed which
brings in a wide range of error. These error bars are
omitted in Figs. 1 and 2 and will not be taken into account in
the following analysis. Thus, Figs. 1 and 2 show only the mean
values for the Ar-37 production rate for the various
individual experiments. The data are unevenly spaced and the
two data sets contain only low numbers of data. Recently, the Ar-37 
production rate data, shown in Fig.1, have been reviewed, taking into 
account counter efficiencies, chemical yields, and background effects 
\cite{Davi96}. The new analysis has yielded an overall rate 
approximately ten percent higher than the earlier analysis. The time 
distribution of the data in Figs. 1 and 2 is essentially unchanged. The 
combined maximum likelihood rate for the 99 experiments in Fig. 2 is 
$0.480\pm 0.032$ Ar-37 atoms per day in 615 tons of perchlorethylen or 
$2.55$ SNU $(1 SNU=10^{-36}$ captures/target atom/sec). This experimental 
result, compared with the theoretical predictions of $8$ SNU for the 
Standard Solar Model, has become known as the solar neutrino problem 
\cite{Bahc89}. The following provides the Fourier spectrum 
analysis of the earlier Ar-37 production data \cite{Haub91} and, for 
comparison, of the new data. The analysis shows that the review of the 
earlier data did not change essentially the resulting power spectral density that reveals a strong harmonic content in the data.

To remove some of the high frequency noise from the data in
Figs. 1 and 2, the five-point moving average is calculated,
which adjusts each point to be the average value of
the five points around it. The results in Figs. 3 and 4 are
longer than original by one point. This causes a slight phase
shift in the display.

The fast Fourier transform of a real series results in a
complex series where the second half of the result is the
mirror image of the first half. The power spectral density is
the magnitude squared of the first half of the fast Fourier
transform and factors out the length of the original series.
The power spectral density in Figs. 5 and 6 is useful for comparing the
frequency spectrum of different series and can be thought of
as the power of a series at a particular frequency \cite{Haub91}.

\section{CONCLUSION}
The investigation of the variation of the Ar-37 production rate over time 
in the Homestake solar neutrino experiment reveals several periodicities 
shorter than or almost equal to 11 years. A number of harmonics contained 
in the Ar-37 data set may have a non-random origin. It is a striking fact 
that the power spectrum density did not change significantly in the various 
analysis undertaken over the years while gradually accumulating data in the 
Homestake experiment. It can not be excluded at this point of time that the 
discovered harmonic content of the data reflects solar activity through a 
not yet known physical phenomena, seated in the deep interior of the sun. 
Attention is drawn to the existence of distinct peaks in the power spectrum 
density of the Ar-37 production rate. Extending and improving of the data 
could be achieved through taking into account the results of the other three 
solar neutrino experiments (Kamiokande, GALLEX, SAGE) if they continue 
operating over a sufficient period of time. In conclusion, the present 
results confirm the existence of periodicities in the new set of Homestake 
data, previously reported for the earlier data.
\clearpage
\begin{figure}[t]
\begin{minipage}[t]{75mm}
\framebox[74mm]{\rule{0mm}{55mm}}
\caption {The observed Ar-37 production rate in the Homestake experiment 
("earlier" data) for runs 18(1970.78)-109 (1990.04)[abscissa: time (years), 
ordinate: Ar-37 production rate (atoms/day)].}
\end{minipage}
\hspace{\fill}
\begin{minipage}[t]{75mm}
\framebox[74mm]{\rule{0mm}{55mm}}
\caption{The observed Ar-37 production rate in the Homestake experiment 
("new data") for runs 18(1970.78)-124 (1992.38)[abscissa: time (years), 
ordinate: Ar-37 production rate (atoms/day)].}
\end{minipage}
\end{figure}

\begin{figure}[hb]
\begin{minipage}[t]{75mm}
\framebox[74mm]{\rule{0mm}{55mm}}
\caption {Five-point moving average of the Ar-37 production rate shown in 
Fig.1.}
\end{minipage}
\hspace{\fill}
\begin{minipage}[t]{75mm}
\framebox[74mm]{\rule{0mm}{55mm}}
\caption{Five-point moving average of the Ar-37 production rate shown in 
Fig.2.}
\end{minipage}
\end{figure}
\clearpage
\begin{figure}[t]
\begin{minipage}[t]{75mm}
\framebox[74mm]{\rule{0mm}{55mm}}
\caption {The power spectral density for the Ar-37 record
(abscissa: frequency (cycles/year), ordinate: power) shown in
Figure 1.}
\end{minipage}
\hspace{\fill}
\begin{minipage}[t]{75mm}
\framebox[74mm]{\rule{0mm}{55mm}}
\caption{The power spectral density for the Ar-37 record
(abscissa: frequency (cycles/year), ordinate: power) shown in
Figure 2.}
\end{minipage}
\end{figure}

Fig. 1 and 2 exhibit periods of high and low Ar-37 production rates indicating that there is an apparent change in the signal. In Figs. 3 and 4 one notes that the five-point moving average in the periods 1977 to 1980 and 1987 to 1990 shows a strongly suppressed signal. The highest peakes in Figs. 5 and 6 appear at about 10.0 yr and 0.5 yr periodicities. Shorter periods of 1.66 yr and 0.71 yr can also be observed.


\begin{thebibliography}{10}

\bibitem{Kirs95} T.A. Kirsten, Ann. N.Y. Acad. Sci. 759 (1995) 1.

\bibitem{Davi92} R. Davis Jr., Ann. N.Y. Acad. Sci. 655 (1992) 209.

\bibitem{Davi96} R. Davis Jr., presentation made at the 188th meeting of 
the American Astronomical Society, 9-13 June 1996, University of Wisconsin, 
Madison, Wisconsin, USA.

\bibitem{Bahc89} J.N. Bahcall, Neutrino Astrophysics, Cambridge University 
Press, Cambridge, 1989.

\bibitem{Haub95} H.J. Haubold and A.M. Mathai, Astrophys. Sp. Sci. 228     
(1995) 113.

\bibitem{Gran96} A.A. Grandpierre, Astr. Astrophys. 308 (1996) 199.

\bibitem{Dick78} R.H. Dicke, Nature 276 (1978) 676.

\bibitem{Rabe89} G.F. Rabey and H.A. Hill, Ann. N.Y. Acad. Sci. 571 (1989) 
594. 

\bibitem{Haub91} H.J. Haubold and J. Beer, in Solar-Terrestrial Variability 
and Global Change (Eds. W. Schroeder and J.P. Legrand), Proceedings of the 
IUGG/IAGA General Assembly, Vienna, 1991.

\bibitem{Mcnu95} R.L. McNutt Jr., Science 270 (1995) 1635.


\end{thebibliography}
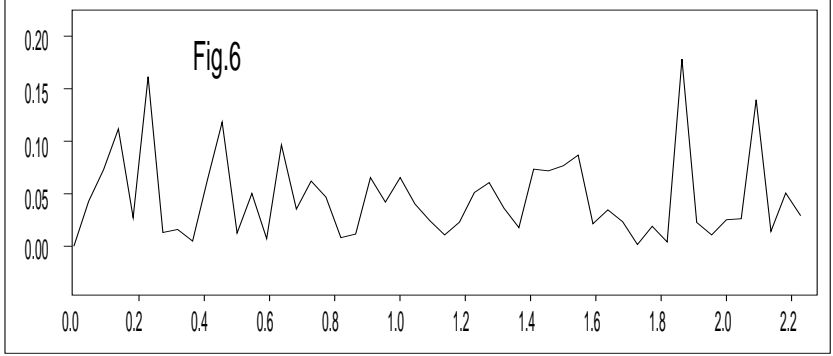
\end{document}